\documentclass[fleqn,usenatbib]{mnras}
\usepackage{newtxtext,newtxmath}
\usepackage[T1]{fontenc}
\usepackage[utf8]{inputenc}
\usepackage{ae,aecompl}

\usepackage{graphicx}	
\usepackage{amsmath}	
\usepackage{amssymb}	

\title{Galaxy shape measurement with convolutional neural networks}

\author[D. Ribli et al.]{
Dezső Ribli$^{1}$\thanks{E-mail: dkrib@caesar.elte.hu (DR)},
László Dobos$^{1}$
and István Csabai$^{1}$
\\
$^{1}$Eötvös Loránd University, Institute of Physics, Budapest, Hungary}

\date{Accepted XXX. Received YYY; in original form ZZZ}

\pubyear{2019}

\begin{document}
\label{firstpage}
\pagerange{\pageref{firstpage}--\pageref{lastpage}}
\maketitle

\begin{abstract}

We present our results from training and evaluating a convolutional neural network (CNN) to predict galaxy shapes from wide-field survey images of the first data release of the Dark Energy Survey (DES  DR1).
We use conventional shape measurements as ``ground truth''  from an overlapping, deeper survey with less sky coverage, the Canada-France Hawaii Telescope Lensing Survey (CFHTLenS).
%
We demonstrate that CNN predictions from single band DES images reproduce the results of CFHTLenS at bright magnitudes and show higher correlation with CFHTLenS at fainter magnitudes than maximum likelihood model fitting estimates in the  DES Y1 \textsc{im3shape} catalogue.
%
Prediction of shape parameters with a CNN is also extremely fast, it takes only 0.2 milliseconds per galaxy, improving more than 4 orders of magnitudes over forward model fitting.
%
The CNN can also accurately predict shapes when using multiple images of the same galaxy, even in different color bands, with no additional computational overhead.
The CNN is again more precise for faint objects, and the advantage of the CNN is more pronounced for blue galaxies than red ones when compared to the DES Y1 \textsc{metacalibration} catalogue, which fits a single Gaussian profile using $riz$ band images.
We demonstrate that CNN shape predictions within the  \textsc{metacalibration} self-calibrating framework yield shear estimates with negligible multiplicative bias, $m<10^{-3}$, and no significant PSF leakage.
%
Our proposed setup is applicable to current and next generation weak lensing surveys where higher quality ``ground truth'' shapes can be measured in dedicated deep fields.

\end{abstract}




\section{Introduction}

Light from distant galaxies is deflected by the tidal fields of inhomogeneous matter along the line of sight, distorting the shapes of sources, a phenomenon called gravitation lensing.
Cosmic shear is the weak gravitational lensing effect arising from no obvious foreground mass apart from the large scale structure of the Universe.
Detecting the effects of the foreground mass density field on distant galaxy images allows us to  indirectly map the distribution of the elusive, and apparently very abundant dark matter.
Furthermore, characterising matter distribution at different redshifts through the lensing signal offers a unique window to the evolution of the dark energy dominated late Universe, complementary to other observations (see \cite{mandelbaum_weak_2018,kilbinger_cosmology_2015} for recent reviews).

%
When the lensing effect of foreground matter is strong enough, it can dramatically alter the appearance of background sources, making the lensing signal detectable even for individual galaxies.
The large scale structure, however, only distorts the shapes of background objects with a small, approximately linear shear, rendering the lensing signal indistinguishable from intrinsic shape variations of lensed galaxies.
The spatial correlations in the apparent shapes of galaxies introduced by the large scale structure make cosmic shear statistically measurable from an ensemble of galaxies.
Since the first detection of cosmic shear about two decades ago \citep{bacon2000detection,kaiser2000large,van2000detection,wittman2000detection}  weak lensing measurements have matured into a potent probe of cosmology.
Weak lensing survey volumes increased  constantly,  with the COSMOS field and Hubble Space Telescope images \citep{schrabback2010evidence} followed by
CFHTLenS \footnote{http://www.cfhtlens.org}  which was the first major weak lensing survey pushing the number of galaxies with high-quality shape measurements to the millions by covering an area of 154 square degrees at a resolved galaxy density of 17 per square arc minute.
Shapes of hundreds of millions  of galaxies are being measured by ongoing weak lensing surveys, such as the Dark Energy Survey (DES) \footnote{https://www.darkenergysurvey.org}, the Kilo-Degree Survey (KiDS) \footnote{http://kids.strw.leidenuniv.nl}, and the Subaru Hyper Suprime-Cam Survey (HSC) \footnote{https://hsc.mtk.nao.ac.jp/ssp/}, whereas billions more will be observed by large surveys that are under construction, such as the Large Synoptic Survey Telescope (LSST) \footnote{http://www.lsst.org/}, or in the development phase, such as the Euclid mission \footnote{http://sci.esa.int/euclid/} or the Wild Field Infrared Survey (WFIRST) \footnote{http://wfirst.gsfc.nasa.gov/}.

Cosmic shear only distorts the shapes of galaxies at percent levels \citep{kilbinger_cosmology_2015}, and the signal is dominated by noise from intrinsic shape variation of galaxies, atmospheric and instrumental distortions. 
The bulk of the cosmic shear signal is carried by shapes of small and faint galaxies typically those with sizes of a few arcsecs and ~24 magnitudes  \citep{kilbinger_cosmology_2015}.
Weak lensing surveys, therefore, need to accurately estimate the shapes of galaxies which only cover a few pixels and have very low signal to noise ratios.
%
Shape measurements are further complicated by the smearing of the original galaxy images by atmospheric and instrumental point spread functions (PSF).
The PSF itself has a coherent spatially and temporally variable anisotropy mimicking the effect of cosmic shear, therefore PSF anisotropy must be carefully measured and corrected for precise shear estimation \citep{miller_bayesian_2013}.

Popular shape measurement algorithms fall in two main categories, the first approach attempts to directly measure the shapes of galaxies through second-order moments of the surface brightness profile \citep{kaiser1995method}.
%
Another approach, dominantly used in recent weak lensing surveys, uses forward model fitting, where parametric surface brightness model profiles are convolved with the PSF and compared to measurements \citep{kuijken1999weak, miller_bayesian_2007, kitching_bayesian_2008,miller_bayesian_2013,kuijken2015gravitational,hildebrandt_kids-450:_2017,conti_calibration_2017,zuntz_dark_2018}.
Model fitting is preferred over measuring moments due to the efficient handling of the PSF and the joint fit over multiple exposures \citep{heymans2012cfhtlens} or colours \citep{zuntz_dark_2018}.
In model fitting approaches it is convenient to parametrise the shapes of galaxies with the shear style complex ellipticity $\boldsymbol{\epsilon}$,  which is an unbiased estimator of weak shear.

\begin{equation}
\epsilon_1 + i\, \epsilon_2 = \frac{a-b}{a+b} e^{ i\, 2 \theta}
\end{equation}

, where $a$ is the length of the semi-major axis, $b$ is the length the semi-minor axis and $\theta$ is the position angle of an ellipse.

The redshifts of galaxies also need to be measured for optimal extraction of the cosmological signal, however, this is not possible with spectroscopy due to the vast number of galaxies used, and weak lensing surveys need to rely on photometric redshift estimates from multi-band observations.
Shapes of galaxies are highly correlated in different colors and information from multiple bands can be combined to improve shape measurements \citep{jarvis_combining_2008}.
The feasibility of this approach was demonstrated in one of the DES Y1 shape catalogues \citep{zuntz_dark_2018}.

Pixel values of a galaxy image have complex non-linear relationship with ellipticity parameters and the likelihood surface is skewed towards zero ellipticity and towards the ellipticity of the PSF.
This effect introduces a bias in the maximum likelihood estimate or the expected value of the shape parameters, which becomes significant for noisy galaxies, hence it is called the ``noise bias'' \citep{bernstein_shapes_2002,kitching_great10:_2012,melchior2012means,refregier2012noise,kacprzak2012measurement,miller_bayesian_2013}.
Following \cite{heymans_shear_2006}, bias in ensemble shear measurements is usually parametrized with a linear function.

\begin{equation}
\langle \epsilon_i^{obs} \rangle = (1+m)\,  g_i + c + \alpha  \epsilon_i^{PSF} 
\end{equation}

, $i$ denotes the component of the ellipticity (1,2), $g$ is the true cosmic shear, the intercept $c$ is called the additive bias, the excess slope $m$ is the multiplicative bias and  $\alpha$ is the PSF leakage.
The additive bias and the PSF leakage are strongly related and they are corrected using measurement data \citep{heymans2012cfhtlens}, however, the multiplicative term is generally mitigated with simulations.

The multiplicative bias strongly depends on the observed S/N and size of galaxies and the precise relationship is identified through estimating the properties of millions of galaxy images simulated with sophisticated tools \citep{rowe_galsim:_2014} and known shape parameters \citep{miller_bayesian_2013,jarvis_science_2016,conti_calibration_2017,zuntz_dark_2018,pujol2019highly}.
These simulations need to be specifically tailored to the details of surveys, and the calibration process has to be conducted for every single survey or data release.
After the precise nature of the bias is established, empirical corrections of measured galaxy ellipticities are applied to negate the bias.

The DES Y1 shape catalogue \citep{zuntz_dark_2018} employed an innovative approach called \textsc{metacalibration}, which only uses observational data and post measurement shear operations and convolutions to correct for noise bias \citep{huff_metacalibration:_2017, sheldon_practical_2017}.
\textsc{metacalibration} offers a principled shear calibration framework without large numbers of calibration factors derived from simulations, with sufficiently small multiplicative bias even for large future weak lensing surveys \citep{sheldon_practical_2017}.
The PSF anisotropies are handled during the deconvolution step in \textsc{metacalibration}, which allows the calibration of any well-behaving estimator \citep{sheldon_practical_2017}, such as adaptive moments \citep{bernstein_shapes_2002}, not only model fitting with forward convolutions.

Apart from noise bias, selection biases also appear when galaxies are preferentially detected depending on the alignment of their shapes, the anisotropy of the PSF and lensing shear \citep{kaiser2000new, hirata2003shear}.
Selection bias may also affect detected galaxies due to the fact that the ellipticity likelihood surface is narrower for galaxies if the intrinsic and PSF ellipticities are aligned.
Methods which assign an inverse-variance weight to shape measurements based on the width of the likelihood surface \citep{miller_bayesian_2013, conti_calibration_2017}, favour galaxies parallel to the PSF, creating a selection bias in the shear.

Shape measurement approaches which assume simplified galaxy surface brightness profiles potentially suffer from 'model bias' if they are not able to capture the shapes of complex galaxies \citep{voigt2010limitations,melchior2010limitations}.
Model bias is expected to have only a  minor contribution in ground-based observations \citep{miller_bayesian_2013,mandelbaum_great3_2015}, and it was explored in detail for model fitting shape measurements  \citep{zuntz2013im3shape,kacprzak2014sersic}.

Biases in the shape measurements could systematically alter cosmological parameters inferred from weak lensing measurements and the community understandably paid particular attention to the question of systematic biases, also in the form of organized challenges
 \citep{heymans_shear_2006,massey_shear_2007,bridle_great08_2009,bridle_great08:_2010,kitching_great10_2010,kitching_great10:_2012,mandelbaum_great3_2014,mandelbaum_great3_2015}.
However, shape measurements not only need to be unbiased but also they need to have small variance in order to reduce statistical uncertainty, a quantity often expressed in terms of an effective galaxy surface density \citep{heymans2012cfhtlens,chang_effective_2013}.
Large efforts are ongoing and planned (HSC, LSST, Euclid, WFIRST) to conduct very deep surveys which aim to radically increase the galaxy surface density and the statistical power of measurements \citep{takada2010subaru,chang_effective_2013,amendola2018cosmology}.
Increased galaxy densities enable more precise measurements of cosmic shear at high angular resolutions, allowing the extraction of significant non-Gaussian information, which can further empower weak lensing measurements \citep{ribli_weak_2019}.

The variance of shape measurements received much less attention than their bias, probably because one might think that model fitting with forward convolution is an optimal method.
However, noise in galaxy shape measurements is known to be not a completely independent shot noise \citep{mandelbaum_great3_2015} and model fitting approaches can only work with simplified parametric surface brightness profiles \citep{miller_bayesian_2013,zuntz_dark_2018}, which may be suboptimal for faint irregular galaxies, abundant in weak lensing surveys.
Systematic biases in current surveys are significantly smaller than statistical uncertainties \citep{mandelbaum_great3_2015,zuntz_dark_2018,conti_calibration_2017}, therefore shape measurements with smaller dispersion could further reduce the uncertainty of underlying cosmological parameters.

Maximum likelihood model fitting with forward convolutions takes approximately 1.6 seconds per galaxy per image \citep{zuntz2013im3shape}, consuming 200,000 CPU hours for the DES Y1 \textsc{im3shape} shape catalogue.
The final DES survey is expected to include an order of magnitude more galaxies, and near future surveys such as LSST and Euclid will again grow an order of magnitude over that number.
Including multiple colors in the fitting process or handling approximately 1000 individual observations per galaxy in LSST further boost the computational burden of maximum likelihood model fitting.
In order to handle the large amount of data expected in future surveys, extremely fast shape estimators are needed which can handle multi-epoch data efficiently.

This work focuses on reducing the statistical dispersion of shape estimates, via convolutional neural networks (CNN) which recently revolutionized the field of computer vision and reached human level accuracy in image classification \citep{he2015delving}.
CNNs are sophisticated machine learning methods which are able to learn from a large number of labelled images, and apart from image classification they also excel in image deconvolution or "super-resolution" in the presence of noise for everyday images \citep{xu2014deep} or microscopy \citep{wang2019deep}.
We construct a large labelled dataset using images from a wide survey, DES DR1 \citep{abbott_dark_2018,morganson_dark_2018}, and high quality galaxy shape measurements from a significantly deeper survey with overlapping footprint, CFHTLenS \citep{erben_cfhtlens:_2013,heymans_cfhtlens_2013,hildebrandt2012cfhtlens,miller_bayesian_2013}.
The CNN is trained to predict ellipticities of galaxies measured by the deeper survey using images of the wide survey as inputs.

CNNs evaluate an image with a single forward pass though their layers, while making use of massively parallel computer hardware (GPU, TPU) which allow sub-millisecond execution times per galaxy, orders of magnitude faster than model-fitting approaches.
The extreme speed of CNN based estimation alone warrants the exploration of this approach, as it may be able to handle the expected data surge and exploit all the information available at the same time in near future large weak lensing surveys.

Multiple studies explored recently machine learning and CNNs in problems related to weak gravitational lensing or the estimation of galaxy properties
\cite{dieleman2015rotation} used custom rotation invariant CNNs for galaxy morphology prediction using ground truth values determined by citizen scientist in the Galaxy Zoo project \citep{lintott2008galaxy}.
CNNs were used to infer the morphological parameter of simulated galaxies \citep{tuccillo_deep_2016}, with an improved catalogue released for SDSS \citep{dominguez2018improving}.
\cite{herbel_fast_2018} use CNNs for the estimation of PSF shape parameters to enable fast PSF modelling.
\cite{springer2018weak} train CNNs to directly predict relatively strong simulated shears in large, resolved galaxy images in the context of galaxy cluster lensing.
Most recently \cite{tewes2019weak} train neural networks to calibrate shear measurements using biased shape estimators such as second order moments and various galaxy parameters as inputs.
The approach used by \cite{tewes2019weak} is complementary to our work and it could use the output of a shape estimator CNN to achieve potentially more accurate shear estimators with small bias.

The outline of the paper is the following: in \S \ref{dataandmethods} we describe the dataset used for training and testing, the details of the CNN architecture and the training process.
In \S \ref{rbandresults} we train a shape estimator CNN on single band images and compare it to the DES Y1 \textsc{im3shape} catalogue depending of various measured factors.
In \S \ref{rizbandresults} we train the CNN on multiple band images and compare it to the DES Y1 \textsc{metacalibration} catalogue.
In \S \ref{size} we determine the necessary training dataset size.
In \S \ref{meta}  we test the biases of the CNN shape estimator nested in \textsc{metacalibration} using a large suit of simulated galaxies, and finally we discuss the results in \S \ref{discuss}.

\section{Data and methods}
\label{dataandmethods}
\subsection{Observational data}
\label{data}

\begin{figure}    
  \includegraphics[width=1\columnwidth]{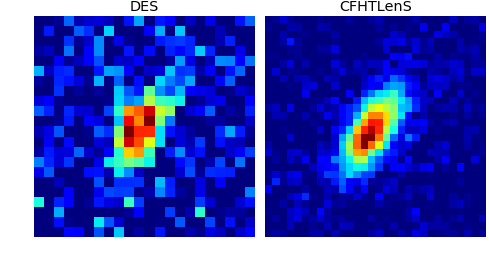}
  \includegraphics[width=1\columnwidth]{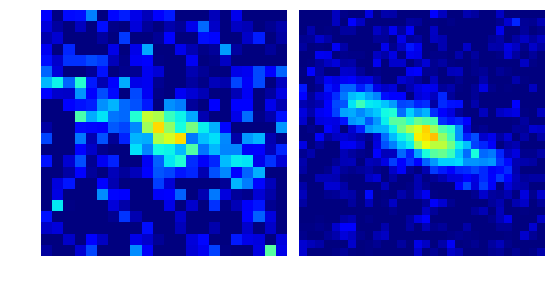}
  \includegraphics[width=1\columnwidth]{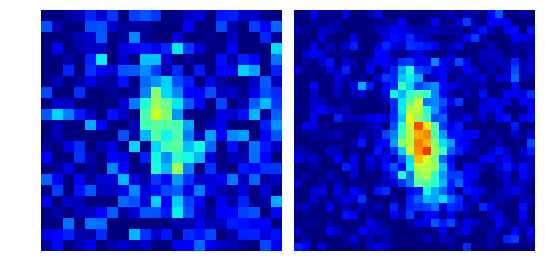}
    \caption{Example galaxies on DES (left) and the deeper CFHTLenS survey(right) images in the $i$-band. CFHTLenS images have higher S/N and they allow more accurate shape measurements of the same galaxies. The galaxies have $i$-band magnitudes $(22.05, \, 22.09, \, 22.18)$, top to bottom.}
    \label{fig:desvscfht}
\end{figure}

We select galaxies from DES Y1 shape catalogues which are also in the overlapping W4 region of CFHTLenS.  
The DES Y1 shape catalogues (\textsc{im3shape},\textsc{Metacalibration}) \citep{zuntz_dark_2018}, are cross matched with the CFHTLenS shape catalogue \citep{erben_cfhtlens:_2013} using Skyquery, a probabilistic join engine for cross-identification of multiple astronomical databases \citep{dobos2012skyquery}.
We only use galaxies with confident CFHTLenS shape measurements, specifically, where the inverse variance weight is larger than $14$.
The final dataset contains  $\approx 1.4\times10^5$ galaxies.

The corresponding DES DR1 stacked multi epoch images \citep{abbott_dark_2018} were downloaded for each galaxy and 48 pixel sized postage stamps were cut out in each of the $'grizY'$ DES bands.
Stacked multi epoch images are considered sub-optimal for shear measurement and were recently replaced by joint fitting of multiple single exposures for the analysis of weak lensing surveys \citep{heymans2012cfhtlens,miller_bayesian_2013,kuijken2015gravitational,conti_calibration_2017,hildebrandt_kids-450:_2017,zuntz_dark_2018}, however, these images are suitable for evaluation of the dispersion of shape estimates.
Potential biases associated with dithered stacked images due to PSF discontinuities and interpolation \citep{miller_bayesian_2013}  are not important here as we do not attempt to go further than shapes of individual galaxies, and we do not calculate ensemble shear estimates.
The depth of the Y1 and the DR1 releases of DES are similar in the part of the equatorial region used in this study, as Y2 and Y3 operations concentrated on other regions of the sky \citep{diehl2016dark}.

We define highly confident ellipticity parameters measured with \textit{lensfit} using the CFHTLenS $i$ band images \citep{miller_bayesian_2013} as the 'ground truth' labels to train our neural network.
The two surveys use telescopes with similar mirror sizes, and CFHTLens $i$ band data was integrated for $4300$ seconds, while the DES DR1 data has $450$ second integrated exposure in each of the $griz$ bands.
The CFHTLenS survey is approximately 1 magnitude deeper than DES DR1 \citep{erben_cfhtlens:_2013,abbott_dark_2018}, which allows much more precise shape measurements for faint galaxies.
The difference in image quality for faint galaxies is demonstrated visually with examples on [Fig. \ref{fig:desvscfht}].

\begin{figure}   \includegraphics[width=1\columnwidth]{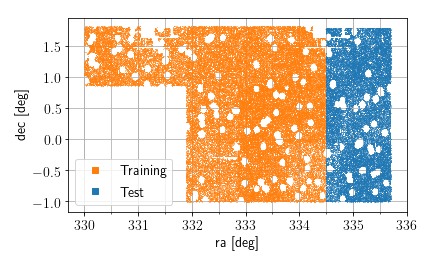}
    \caption{We use non-overlapping training and test regions to evaluate generalisation to new regions in the sky. }
    \label{fig:split}
\end{figure}

We split the dataset into a 70\% training and a 30\% test set based on the position on the sky [Fig. \ref{fig:split}], in order to evaluate the capability of the CNN to generalize to a new region, other than the one used for training.
Generalization to other regions of the sky is essential in our proposed scheme where the network is trained on dedicated deep fields of surveys but used for estimation on the whole wide field survey.

\subsection{Evaluation}

For calibration purposes, the 'accuracy' of shape measurement algorithms is often characterized by the values of multiplicative and additive bias.
However, in this study we set out to reduce the variance, the statistical dispersion of shape measurements, therefore we choose a different metric which reflects the covariance of ground truth and estimated ellipticities: the Pearson correlation coefficient.

\begin{equation}
\rho(x,y) = \frac{cov(x,y)}{\sigma_{x} \, \sigma{y}}
\end{equation}

We do acknowledge the immense importance of bias correction, which must be thoroughly performed for credible shear measurements, and we demonstrate that CNN shape estimates have sufficiently small bias for future large surveys when implemented in the  \textsc{metacalibration} framework.
We also expect that CNN shape estimates can just as well be calibrated using a large number of realistic image simulations as any other well-behaving shape estimator.
Additional parameters for calibration, such as signal to noise ratio and size could also be predicted by the CNN, similarly to shape, or they could be derived from other tools such as SExtractor \citep{bertin96sex}.
For galaxy level comparisons performed in the present study, bias corrections are negligible compared to the intrinsic dispersion of galaxy ellipticities.

\subsection{Convolutional neural network, training and testing}
\label{cnn}

We design a custom CNN architecture specifically for the task, building on successful image classification CNNs \citep{krizhevsky2012imagenet,simonyan2014very,szegedy2015going,he2016deep,redmon2017yolo9000}.
The network consists of subsequent sliding window filter matching operations, called convolutional layers, which can powerfully express the translation invariance of image data.
The width of convolutional filters is $3\times3$ pixels for the majority of layers, as this filter width makes most effective use of parameters  \citep{simonyan2014very}.
Starting from the fifth layer these $3\times3$  convolutional layers are preceded by a bottleneck $1\times1$ ``convolutional'' layer, which compresses input representations to half the number of filters to save computation time and reduce the number of overall parameters.
These bottleneck layers are found in almost every single modern CNN architecture \citep{szegedy2015going,he2016deep,redmon2017yolo9000}.

\begin{table}
\centering
\begin{tabular}{r|l|r}
\# & Layers & Output size \\\hline
1 & Convolution ($3 \times 3 $)    & $48 \times 48 \times 16 $ \\
2 & Convolution ($3 \times 3 $)    & $48 \times 48 \times 16 $ \\
3 & Convolution ($3 \times 3 $)    & $24 \times 24 \times 32 $ \\
4 & Convolution ($3 \times 3 $)    & $24 \times 24  \times 32$ \\
5 & Convolution ($3 \times 3 $)    & $12 \times 12  \times 64 $ \\
6 & Convolution ($1 \times 1 $)    & $12 \times 12 \times 32 $ \\
7 & Convolution ($3 \times 3 $)    & $12 \times 12 \times 64 $ \\
8 & Convolution ($3 \times 3 $)    & $6 \times 6  \times 128 $ \\
9 & Convolution ($1 \times 1 $)    & $6 \times 6   \times 64 $ \\
10 & Convolution ($3 \times 3 $)    & $6 \times 6 \times 128 $ \\
11 & Convolution ($3 \times 3 $)    & $3 \times 3 \times 256 $ \\
12 & Convolution ($1 \times 1 $)    & $3 \times 3 \times 128 $ \\
13 & Convolution ($3 \times 3 $)    & $3 \times 3  \times 256 $ \\
- &  Average Pooling  ($3 \times 3$) &  $1 \times 1 \times 256 $ \\
14 & Dense  & 2 \\
\end{tabular}
\caption{Neural network architecture. Number of trainable parameters: 837,586 . }
\label{tab:nn}
\end{table}

Each convolutional layer, except for the last one, is followed by batch normalization \citep{ioffe2015batch}, which rescales activations in the previous layer in order to stabilize and facilitate training.
Batch normalizations are followed by non-linear activation functions called Recitified Linear Units (ReLU), which take the form $max(0,x)$.
The flat part creates non-linearity, which allows the neural network to approximate complex non-linear functions, while the linear region provides stable, non-vanishing gradients when propagated through very deep networks  \citep{krizhevsky2012imagenet}.
After every 2-3 convolutional layers, the representations are spatially downsampled by only evaluating the convolutional filters at every second position.
Downsampling helps to aggregate localized lower level information into higher semantic levels with lower spatial resolution.
We start with $16$ filters in the first layer and double the number of filters after each spatial downsampling, which is a common practice in deep CNNs.
We find that doubling or quadrupling the number of filters in every layer slightly improves the accuracy of the network, however, we settle with 16 filters to avoid additional runtime.
When the spatial extent of convolutional layers becomes very small ($3\times3$) we average all activations along spatial dimensions, creating a single one-dimensional representation.
Finally, a linear layer with $2$ outputs predicts the ellipticity components of the galaxy.
The outline of the neural network is detailed in [Tab. \ref{tab:nn}].

The inputs to the neural network are $48\times48$ pixel postage stamp images with pixel values rescaled by the means and r.m.s. values of the sky measured with SExtractor \citep{bertin96sex}.
When training with multi-band images, different colors are stacked as different input channels similarly to RGB images.
We do not incorporate pixel weights or masks into out inputs, however, we note that it would be straightforward to simply stack these as additional channels of the input image.
During training, we augment the dataset with random horizontal and vertical flips and random transpositions to combat overfitting, ground truth ellipticities of galaxies are transformed accordingly during augmentation.
The augmentation scheme creates an $8\times$ larger dataset, however, the new examples are not independent.
The applied transformations enable the neural network to learn additional symmetries of the dataset alongside translation invariance represented with convolutional operations.
An interesting branch of research efforts attempt to create neural networks with built-in representation for rotational symmetries \citep{cohen2016group,kondor2018clebsch}, however, these works have not completely matured yet.
During testing, we evaluate each galaxy and its 90 degrees rotated version, which reduces the errors by only approximately 1 \%.
Evaluating a rotated version doubles execution time and results in only a modest improvement in accuracy, however,  we prefer to apply this augmentation because it potentially mitigates additive bias due to its symmetry.
We do not predict on the other 6 flipped and transposed versions of the image, because we find only very small improvements from additional test time augmentation,  and it makes inference significantly slower.

\begin{figure*}  
 \includegraphics[width=2\columnwidth]{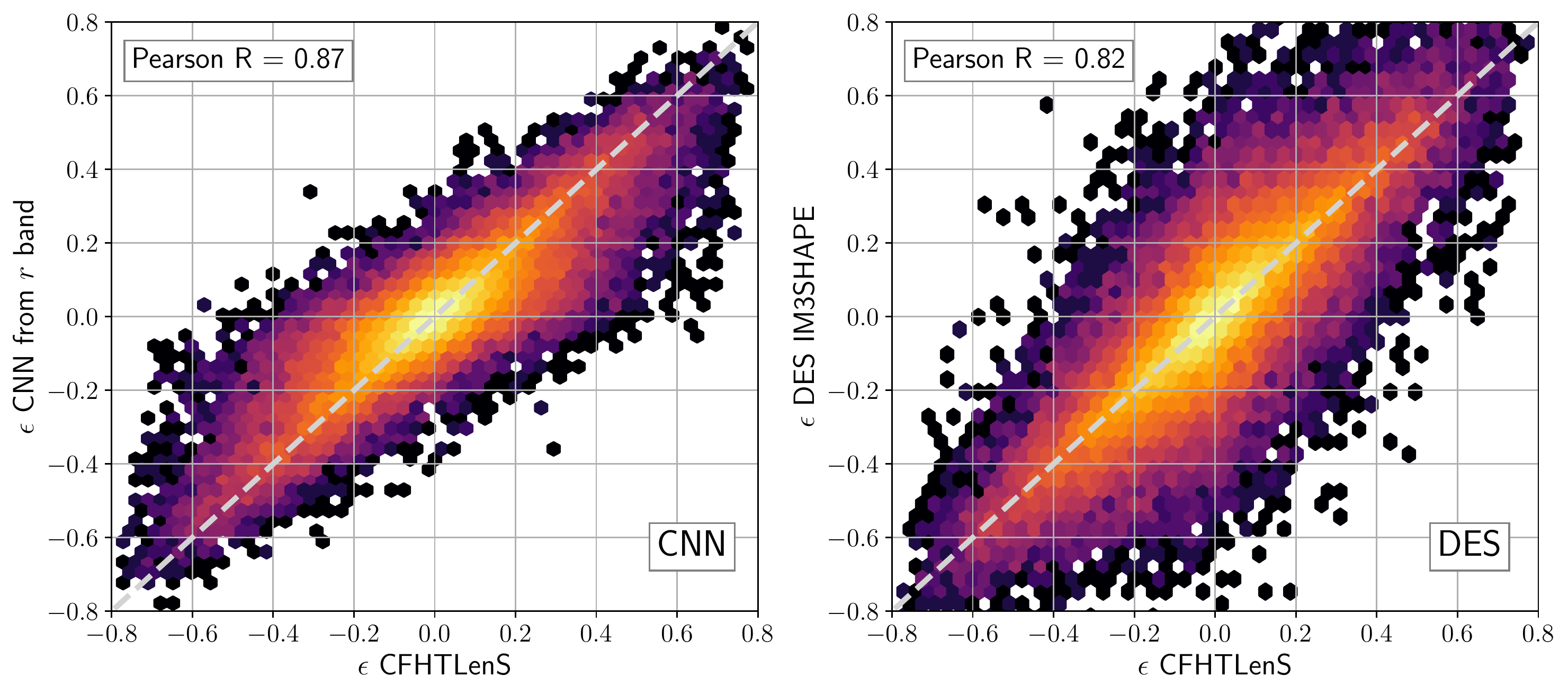}
     \caption{The CNN estimates galaxy shapes with smaller dispersion than the DES \textsc{im3shape} catalogue. The heatmap shows joint distributions of 'ground truth  CFHTLenS shapes  and estimates  by the CNN (left) and the DES shape catalog (right) for the 40\% of galaxies which were selected for cosmology in the DES \textsc{im3shape} catalog.}
    \label{fig:truevsestimate}
\end{figure*}

We train the neural network with a minibatch size of 512  for 40 complete iterations on the training dataset, called an epoch.
The training dataset is reshuffled before each traversal to create a varying composition of minibatches in each iteration, which has a regularization effect when batch normalizations are used \citep{ioffe2015batch}.
We use stochastic gradient descent optimisation with an initial learning rate of $0.005$ and we decrease the learning rate tenfold after the 20th and the 30th epoch to enable convergence by the end of the training schedule.
During training, we minimise the mean absolute error between predicted and target ellipticities.
We do not exhaustively optimize the training process of the neural network by varying hyper-parameters, nor do we attempt to find the optimal CNN architecture.
We find that our base setup is very fast and reasonably accurate, and we delegate fine-tuning of the network details to later works.

Our CNN does not use the known PSF during training or predictions.
The effect of PSF anisotropy is very small compared to the overall variance of shape estimation, and it absolutely does not effect our results, when comparing the shapes of individual galaxies, without ensembling for shear estimation.
Convolutional neural networks are flexible methods and they could be used to produce predictions with low bias and small PSF leakage using special loss functions or similar methods as in  \cite{tewes2019weak}.
However we leave this topic to future work, and in this work we demonstrate that PSF anisotropy and multiplicative bias can be handled by nesting our CNN ellipticity estimator within the \textsc{metcalibration} framework, which can be wrapped around any sufficiently stable ellipticity estimator.
We find that our results are indistinguishable for the two ellipticity components, therefore our results are shown  after concatenating the two components into one ellipticity value.
We make source code for our approach publicly available on github \footnote{https://github.com/riblidezso/shearNN}.

\section{Shape estimation using single band images}
\label{rbandresults}

\begin{figure*}
 \includegraphics[width=2\columnwidth]{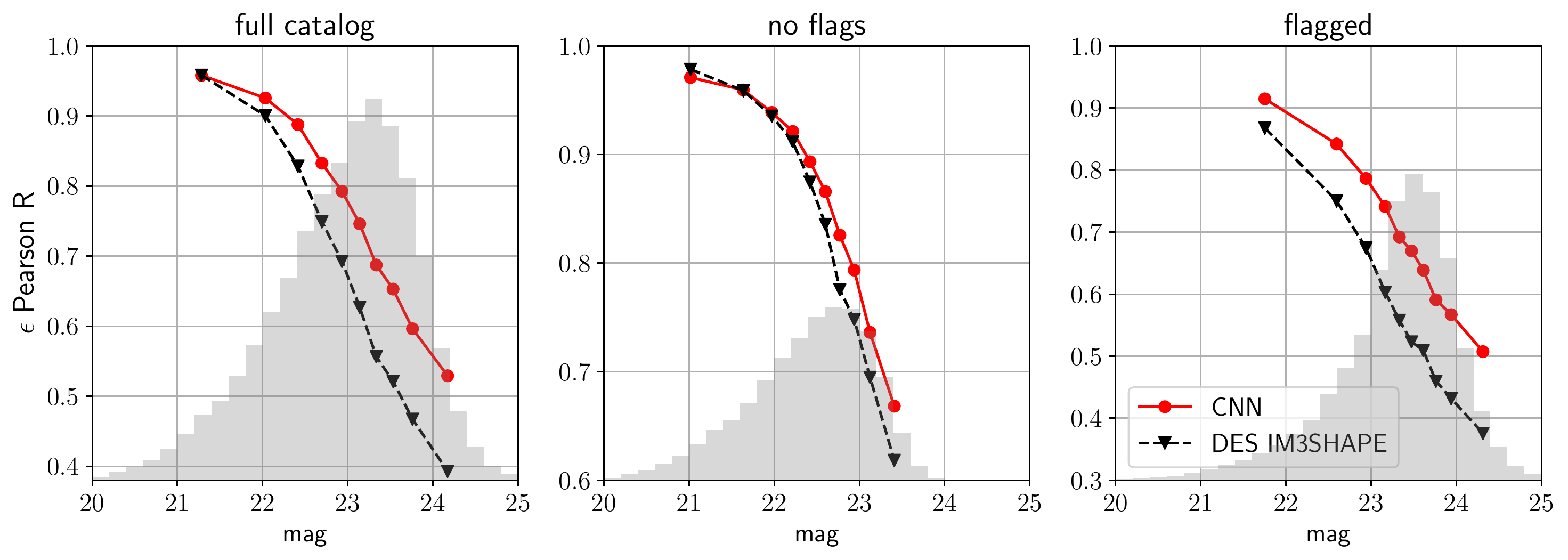}
    \caption{ 
    The shapes of faint galaxies are estimated more accurately by the CNN than the DES \textsc{im3shape} catalog. 
    Pearson correlation between ground truth CFHTLenS and estimated ellipticities are shown depending on the $r$ band magnitude. The histogram of galaxies are shown in light grey.
    On the 40\% subset selected for cosmology, \textsc{im3shape} gives slightly more accurate estimates for bright galaxies, but the CNN is more accurate on much more common faint galaxies (\textbf{center}).
    On the subset excluded from cosmology, mainly due to small size and low brightness (\textbf{right}), the CNN significantly outperforms the DES \textsc{im3shape} catalogue.}
    \label{fig:ered_corr_mag}
\end{figure*}

First, we train and evaluate the CNN shape estimator on only $r$ band stacked DES images, enabling fair comparison with the DES \textsc{im3shape} catalogue.
\textsc{im3shape} is a forward fitting algorithm, where galaxies are modelled with a bulge or a disc profile and profiles are convolved with the PSF before being compared to measured images.
The maximum likelihood solutions are found using the Levenberg-Marquardt algorithm \citep{zuntz2013im3shape}.
For the DES Y1  \textsc{im3shape} catalogue, shape measurements were performed on multi epoch  $r$ band images  \citep{jarvis_science_2016}, and the best fitting bulge or disc models were selected.
A large number of galaxies were flagged unfit for cosmology mainly due to their small size and low signal to noise ratio \citep{jarvis_science_2016},  resulting in unreliable shape estimates, which prohibits accurate bias calibration.
The clean sample only includes around 40\% of galaxies detected both in DES and CFHTLens.
A large portion of these excluded galaxies are selected for cosmology in the deeper \textsc{metacalibration} catalogue.
Heatmaps of  true and  estimated values show that the predictions of the CNN show smaller dispersion than the DES  \textsc{im3shape}  catalogue [Fig. \ref{fig:truevsestimate}] for the high-quality sample selected for cosmology.

\begin{figure*} 
 \includegraphics[width=2\columnwidth]{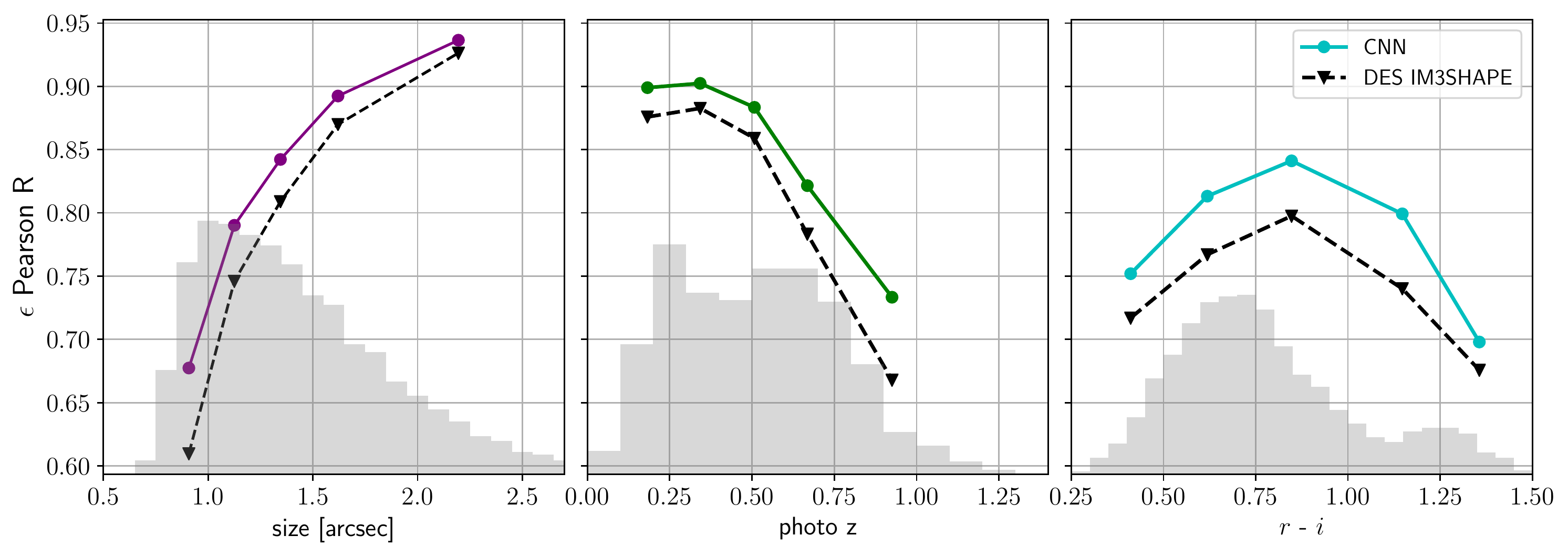}
    \caption{
    The CNN estimates galaxy shapes more accurately than the DES \textsc{im3shape} catalogue regardless of galaxy size (\textbf{left}), redshift  (\textbf{center}) 
 or galaxy $r-i$ color  (\textbf{right}).
 Only galaxies selected for cosmology in the DES \textsc{im3shape} catalogue are shown.
 The $r-i$  color dependence is only shown for galaxies where peak photometric redshift estimate is between $0.6-0.8$.}
    \label{fig:ered_size_z_color}
\end{figure*}

\begin{figure*}  
  \includegraphics[width=0.5\columnwidth]{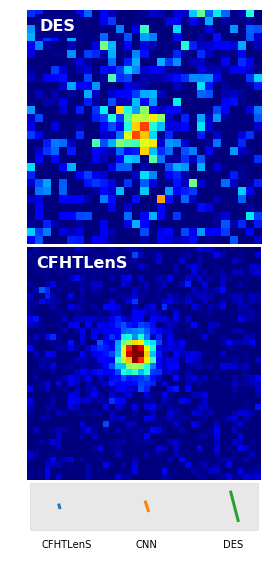}
  \includegraphics[width=0.5\columnwidth]{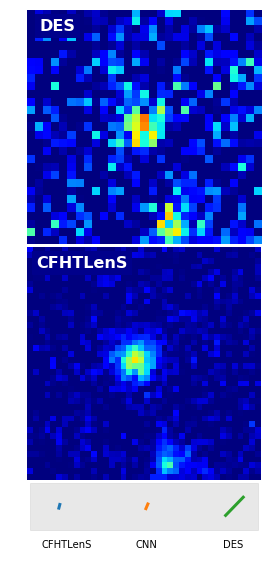}
  \includegraphics[width=0.5\columnwidth]{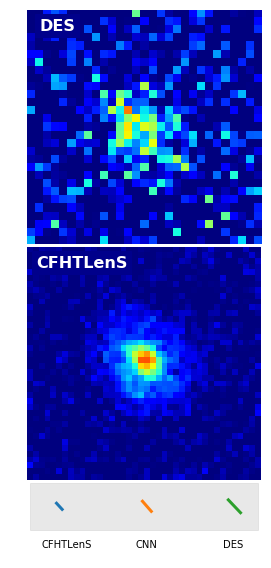}
  \includegraphics[width=0.5\columnwidth]{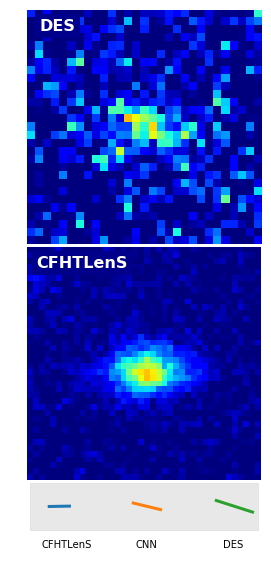}
    \caption{Example galaxy shape measurements and  $r$-band images from DES (top) and $i$-band images CFHTLenS (bottom) for 4 example galaxies. The ellipticity estimates are represented with lines at the bottom of the figures. The height of the grey band represents an ellipticity of 0.8.}
    \label{fig:examples}
\end{figure*}

In order to understand the performance of the CNN shape estimator under various circumstances, we first investigate the accuracy of shape estimates at different magnitudes.
We group galaxies into 10 equal size bins, assign their mean $r$ magnitude to the bin and we calculate the Pearson correlation of estimated and ground truth shapes in each bin [Fig. ] \ref{fig:ered_corr_mag}].
Note, that for these tests we make predictions with the same model, trained on the whole training set, and we only divide galaxies into different bins for the evaluation.
We find that the DES catalogue is just as accurate as the CNN for bright galaxies, and the CNN becomes significantly more precise as noise begins to dominate images.
For clarity, we evaluate the above-mentioned figure for three sets, the full catalogue, the subset selected for cosmology and the one excluded.
Reassuringly, we find that the subset selected for cosmology is much more accurate than the excluded one, and \textsc{im3shape} even outperforms the CNN on the brightest galaxies, and only falls behind for much more abundant faint galaxies.
On the other hand, the CNN vastly outperforms \textsc{im3shape} on the subset of small and faint galaxies which were rightfully excluded from cosmology in the \textsc{im3shape} catalogue.

We also evaluate performance depending on other measured physical properties of galaxies, such as  size, redshift, and color index, to see whether any of these can explain the performance advantage of the CNN.
For these tests, we only use the subset selected for cosmology where \textsc{im3shape} closely matches the performance of the CNN, and galaxies are divided into 5 bins.
In the third evaluation with color index, we selected galaxies with peak photometric redshifts between $0.6-0.8$ according to CFHTLenS photo-z estimates \citep{hildebrandt2012cfhtlens} to restrict our analysis to galaxy colors resulting from different spectral types.
At this redshift, the $4000A$ break lies on the border of  $r$ and $i$ bands, and galaxies clearly show red-blue bimodal distribution.
We selected uniform intervals for the color index bin edges instead of equal size bins in order to sufficiently cover the less populated mode of red galaxies.
We find that the CNN is more accurate than  \textsc{im3shape} regardless of galaxy size, redshift or color index  [Fig. \ref{fig:ered_size_z_color}], however, at large galaxy size, the difference between the methods seems to diminish.
The results on distant, faint and small galaxies indicate that the CNN shape estimator could increase the galaxy density of the survey via allowing fewer galaxies to be excluded than with the  \textsc{im3shape} maximum likelihood estimator.

We manually inspected many galaxies where the CNN is more accurate than \textsc{im3shape} in order to see whether the accuracy difference can be explained by some unusual feature, such as blending or artefacts.
We find no such distinguishing features, except for high noise level and the small galaxy size discussed earlier.
Finally, we show a few examples of galaxies and estimated ellipticities.
We selected two examples where the CNN estimated shapes significantly more accurately than the DES catalog [Fig. \ref{fig:examples}], one where both were correct, and one where both were wrong.
The selected galaxies have high redshifts and large noise and we only selected galaxies where at least 5 different exposures are used when fitting \textsc{im3shape}.
The examples illustrate that galaxy shape estimation is indeed a very hard task for noisy and small examples.

\section{Shape estimation using multi-band images}
\label{rizbandresults}

    \begin{figure*}  
 \includegraphics[width=2\columnwidth]{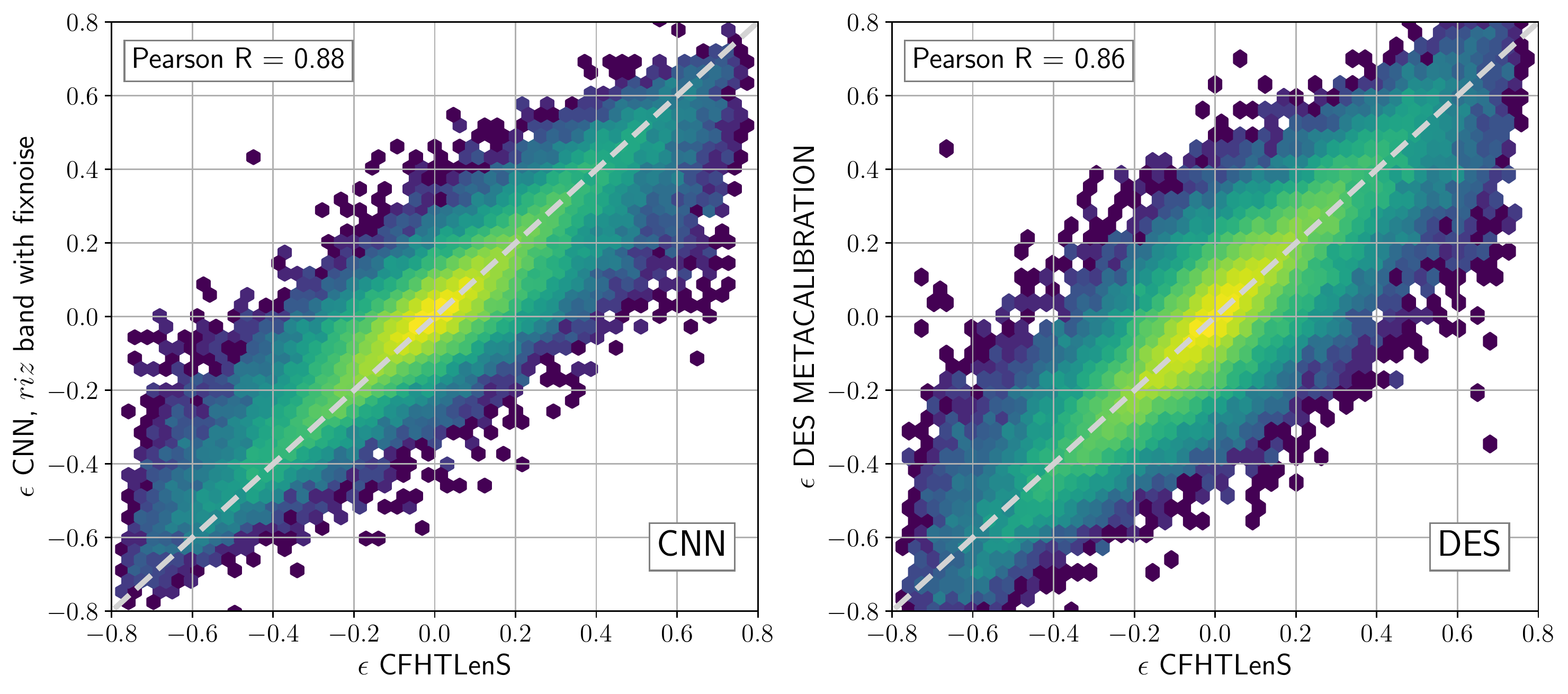}
     \caption{The CNN estimates galaxy shapes with smaller dispersion than the DES \textsc{metacalibration} catalogue.
      The heatmap shows joint distributions of ground truth  CFHTLenS shapes  and estimates  by the CNN (\textbf{left}) and the DES shape catalogue (\textbf{right}) for galaxies which were selected for cosmology in the DES \textsc{metacalibration} catalogue. }
    \label{fig:truevsestimate_meta}
\end{figure*}

\begin{figure}  
 \includegraphics[width=1\columnwidth]{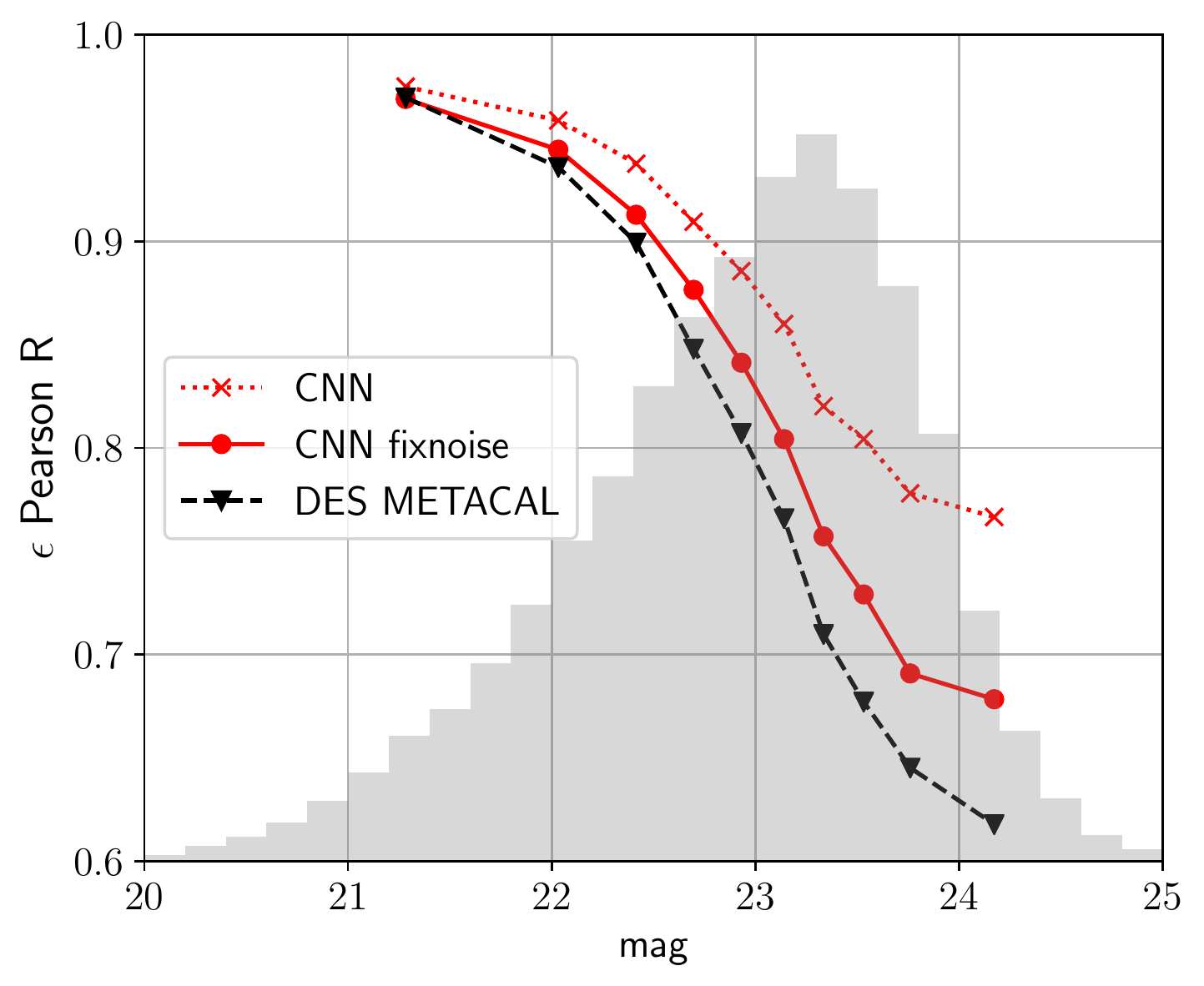}
     \caption{The CNN estimates shapes of faint galaxies more accurately than the DES Y1 \textsc{metacalibration} catalogue.
     The solid line shows the precision of the CNN shape estimator evaluated on images comparable to DES measurements, with additional artificial noise as in \textsc{metacalibration}.
     The dotted line represents CNN shape estimates on original images,  it is added to illustrate the effect of additional noise. 
      Only galaxies selected for cosmology in the DES \textsc{metacalibration} catalogue are shown.}
    \label{fig:ecorr_meta}
\end{figure}

\begin{figure*} 
 \includegraphics[width=2\columnwidth]{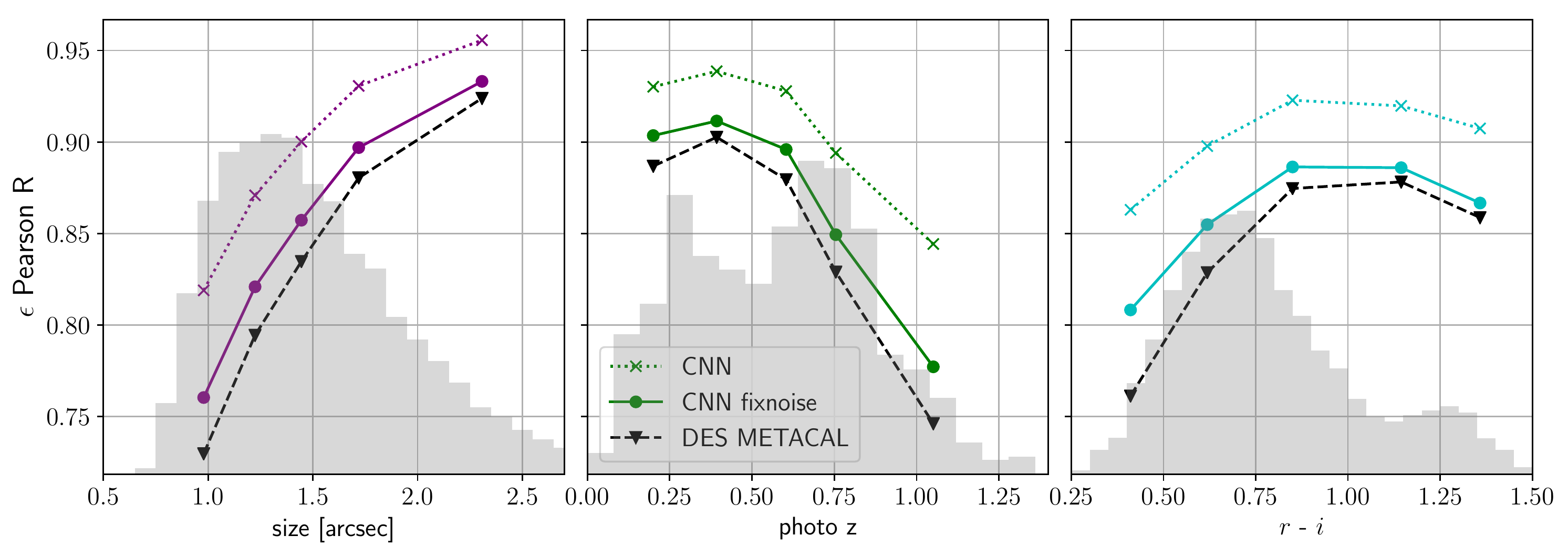}
    \caption{The CNN estimates galaxy shapes more accurately than the DES \textsc{metacalibration} catalogue regardless of galaxy size (\textbf{left})and redshift (\textbf{center}).
    The advantage is larger for blue galaxies than red ones (\textbf{right}). 
  }
    \label{fig:ecorrzcolor_meta}
\end{figure*}

Shapes of galaxies are highly correlated in different colours, therefore different band DES images can be combined to achieve more accurate shape measurements of faint galaxies. 
The most useful colours are $riz$ bands, which are used in the \textsc{metacalibration} catalogue, therefore we also train and evaluate the CNN on images where these 3 bands were stacked in different input channels.
All the other details of the neural network and the training protocol is the same as in the one band case.
Note that additional input channels have practically no effect of the runtime of the CNN, unlike in the case of forward fitting methods, where there is a linear relationship between the number of images in the joint fit and the runtime of estimation.
The first layer in the CNN, which maps from the input channels to 16 filters, is practically negligible compared to the next layer which maps from 16 to 16 filters, regardless of the whether the number of input channels is 1 or 3.
With a much larger number of input channels, the runtime can be dominated by the first layer, and this will inevitably lead to additional computation time, however, with only sub-linear scaling.

We compare the performance of the multi-band CNN to the DES Y1 \textsc{metacalibration}  shape catalogue, which used the 'ngmix' engine \citep{sheldon2015ngmix} to fit a single Gaussian galaxy brightness profile \citep{zuntz_dark_2018}, after forward convolution by the PSF which is also approximated with a single Gaussian.
The simple representation of both galaxy and PSF profiles with a single Gaussian was reported to be necessary for computational efficiency \citep{zuntz_dark_2018}.
Similarly to the \textsc{im3shape} catalogue, fits were performed on many single epoch observations by maximizing joint likelihood.
The ellipticity of the galaxy profiles had a prior strongly centred on zero, which was found to be necessary for stable fits for faint objects.
Note that \textsc{im3shape} does not use such a strong prior on ellipticities which may result in unstable fits for faint galaxies.
This prior results in a strong multiplicative bias ($\approx 0.3$) in raw ellipticity estimates, however, this is entirely corrected with the \textsc{metacalibration} procedure, which we discuss in detail in \S \ref{meta}.
Therefore, we use the ensemble corrected shear estimates for the \textsc{metacalibration} catalogue.
Note that due to the two additional bands, the \textsc{metacalibration} catalogue is significantly more accurate than the \textsc{im3shape} catalogue.

It is important to note that in the \textsc{metacalibration} procedure,  an extra noise field is added to the images before shape measurement.
This noise field has the same r.m.s. as the noise on the original image, effectively reducing the S/N of the image by $\sqrt 2$.
Therefore, in order to perform a fair comparison of the CNN and the \textsc{metacalibration} catalogue we evaluate the CNN on images with an additional noise field with unit variance, called ``fixnoise''.
Predictions on raw images are also shown in order to illustrate the effect of fixnoise.
Note that we train the CNN on raw images, and only add extra noise before predictions.

We repeat the same evaluations as in the case of single-band measurements, using only the subset of galaxies selected for cosmology in the DES Y1  \textsc{metacalibration} catalogue.
In the \textsc{metacalibration} catalogue the only significant exclusion criteria is $S/N>10$ in order to avoid implicit selection effects from source detection.
We find that the CNN estimates shapes of galaxies with smaller dispersion than the DES catalogue when using multi-band images [Fig. \ref{fig:truevsestimate_meta}].
The advantage of the CNN only appears at faint galaxies  [Fig. \ref{fig:ecorr_meta}] and it is independent of the size and the redshift of these galaxies [Fig. \ref{fig:ecorrzcolor_meta}].

Interestingly, we find that the advantage of the CNN is increased on blue galaxies  [Fig. \ref{fig:ecorrzcolor_meta}], while it almost disappears for red ones.
Blue, star-forming galaxies can have complex, irregular shapes and these are apparently much better characterized by the CNN than a single Gaussian fit.
On the other hand,  in the case of relaxed, red, elliptical galaxies a Gaussian profile fits the images almost as well as the predictions of the CNN.
The CNN is not limited by design to use a small set of simple surface brightness profiles, and our results indicate that it is able to learn a useful prior of possible complex galaxy shapes, which gives it a significant advantage over a single Gaussian profile for distant blue galaxies.

\section{The effect of the size of the training dataset}
\label{size}

 \begin{figure} 
 \includegraphics[width=\columnwidth]{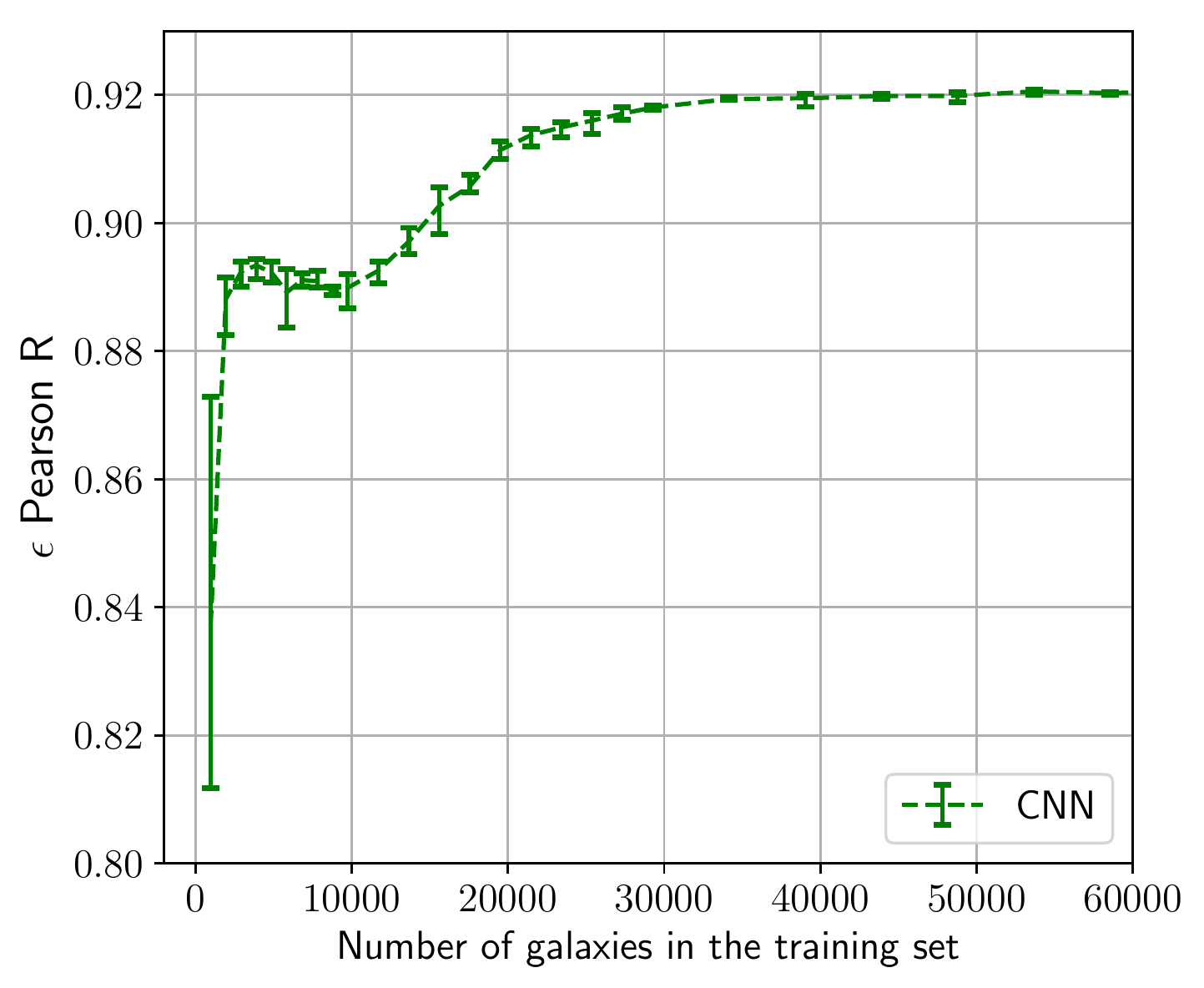}
    \caption{Approximately $4 \times 10^4$ images are needed to reach near maximal accuracy with the CNN using $grizY$ band DES images. 
    Errorbars represent minimum and maximum values for 4 runs with different random training subsets.}
    \label{fig:trainsize}
\end{figure}

Our training dataset contains approximately $10^5$ galaxies, and we investigate the performance depending on the number of galaxies for training the CNN, by randomly excluding galaxies from the training set. 
We maintain the same number of minibatch iterations regardless of training data size, in order to allow convergence for small training data sizes.
For demonstration purposes, we also incorporate  $g$ and $Y$ bands into the input, therefore it is a 5 channel image.
We find that inclusion of the image bands only very slightly improves  performance.
The precision of the CNN plateaus around $4\times 10^4$ galaxies [Fig. \ref{fig:trainsize}], with very little if any improvement for larger training set sizes.
A peculiar flat region appears between $2\times10^3 - 10^4$ galaxies, but this is probably due to sub-optimal training schedule for some training set sizes.
The results indicate that a larger collection of galaxies, such as a simulated dataset, is not expected to significantly improve the dispersion reached here.
Future surveys will easily collect the amount of data necessary to train a CNN shape estimator to reach its full potential.

\section{\textsc{metacalibration}  for the CNN}
\label{meta}
\subsection{The \textsc{metacalibration}  formalism}

\textsc{metacalibration}  is a recently developed innovative self-calibration protocol for weak lensing shear measurements, which calculates necessary corrections using the images themselves instead of deriving them from a large suite of simulated galaxies \citep{huff_metacalibration:_2017,sheldon_practical_2017}.
\textsc{metacalibration} was successfully implemented to produce one of the shape catalogues of the DES Y1 data release \citep{zuntz_dark_2018}.
Without weak lensing, we would measure the unperturbed shape of the galaxy and percent level cosmic shear only very slightly changes the measurement.
For small enough shears the observed shape can be approximated with a linear relationship to shear.

\begin{equation}
\boldsymbol{\epsilon} =\boldsymbol{ \epsilon}|_{\boldsymbol{g}=0}  + \frac{\partial \boldsymbol{\epsilon}}{\partial \boldsymbol{g}} \Bigr|_{\boldsymbol{g}=0} \boldsymbol{g} + O(\boldsymbol{g}^2)
\end{equation}

, where $\boldsymbol{\epsilon} $ is the two component shape estimate and $\boldsymbol{g} $ is the cosmic shear.
The partial derivative of  shape measurement with respect to shear is called response.

The core idea of \textsc{metacalibration} is that the response can be approximated by artificially shearing and remeasuring the image of the galaxy.
Via measuring the shapes of galaxies at different artificial shear values we can calculate the finite derivative of shape measurement with respect to shear.
To correctly mimic the effect of cosmic shear, one must first deconvolve the images with the PSF, apply the known artificial shear and then reconvolve image with a circular synthetic PSF.
It is practical to calculate a two-sided derivative.

\begin{equation}
R_{i,j} = \frac{\epsilon_i (\Delta g_j) - \epsilon_i (-\Delta g_j )}{2\Delta g_j }
\end{equation}

The mean shape over a set of galaxies is the sum of intrinsic shapes and shapes induced by shear.

\begin{equation}
\langle \boldsymbol{\epsilon} \rangle = \langle \boldsymbol{ \epsilon}|_{\boldsymbol{g}=0} \rangle  +  \langle  \boldsymbol{R} \boldsymbol{g}   \rangle + O(\boldsymbol{g}^2)
\end{equation}

The mean shape of galaxies in the absence of shear has zero expectation value for a well-behaving shape estimator.
Assuming constant shear acting on galaxies, the \textsc{metacalibration} shear estimator can be expressed with the mean response and the mean shape over a set of galaxies.

\begin{equation}
\langle \boldsymbol{g} \rangle \approx \langle  \boldsymbol{R} \rangle^{-1}   \langle \boldsymbol{ \epsilon} \rangle  
\end{equation}

For ellipticity measurements, the response is a diagonal matrix, and the formula reduces to an element-wise division.
Note that response is practically equivalent to multiplicative bias, and therefore it is low for faint galaxies, resulting in the implicit down-weighting of  shear measured in these galaxies by the shape estimation algorithm itself as $ \langle \boldsymbol{ \epsilon} \rangle =   \langle  \boldsymbol{R} \boldsymbol{g}   \rangle $.

\subsection{\textsc{fixnoise}  correction }

In the presence of noise, simple \textsc{metacalibration}  breaks down.
The artificial shear is intended to only shear the galaxy, however, it also shears the deconvolved noise field, therefore it fails to accurately mimic cosmic shear. 
The additional sheared correlated noise field results in the underestimation of the response \citep{sheldon_practical_2017}.
In the presence of an asymmetric PSF, another problem also arises.
Apart from the light of the galaxy which was smeared by the PSF originally, the white noise field of the image is also deconvolved by the asymmetric PSF, introducing significant PSF leakage.

An additional noise field, with the same dispersion, but opposite shear statistically  negates the undesired contribution of the original noise field, this method is called ``fixnoise''.
Practically the new noise field needs to undergo the same operations as the images, rotated by  $90\deg$ and added to the manipulated galaxy image. 
The final counter-factual image will be symmetric regarding the direction of PSF asymmetry and artificial shear, therefore the shape measurement will not have PSF leakage and the response can be measured correctly.

\begin{equation}
\widetilde{I} = \widetilde{I_{gal}}  + \widetilde{\eta} ( \Delta \boldsymbol{g}, \boldsymbol{\epsilon}_{PSF})  + \widetilde{\eta_{fix}} ( -\Delta \boldsymbol{g}, -\boldsymbol{\epsilon}_{PSF}) 
\end{equation}

, where $\widetilde{I} $ is the counterfactual image of the galaxy and $\widetilde{\eta}$ is the distorted noise field.
The accuracy of \textsc{metacalibration} with fixnoise correction has been thoroughly verified in simulations by \cite{sheldon_practical_2017} and \cite{zuntz_dark_2018}.

The shape of the galaxy must also be measured on devonvolved and reconvolved images with additional fixnoise, because we measure the response of this particular shape estimator.
The additional noise field reduces the signal to noise ratio of the image by $\sqrt{2}$, reducing the precision of measurements in exchange for principled shear calibration.

\subsection{Simulated galaxies}
\label{simdata}

In principle \textsc{metacalibration} does not need large simulations which precisely match survey data to produce calibration values, however, it is necessary to test the biases of \textsc{metacalibration} with every new shape estimator, pipeline or survey with a large amount of controlled simulation data resembling the true dataset.
We simulate $48 \times 48$ pixel size postage stamp images of $3.5\times10^{8}$ bulge+disk composite galaxies using the galsim python package \citep{rowe2015galsim}, at a scale of $0.263$'' per pixel.
The disk component has an exponential profile and the bulge has a Sérsic profile with $n=4$.
The fraction of flux in the disk component has a uniform distribution between 0 and 1, and the total flux of the galaxies has a truncated Gaussian distribution, the mean flux was 150, the r.m.s. 125, and minimum flux was 100.
We assign uncorrelated ellipticities to the disc and the bulge with Gaussian distributions for the $\epsilon_{1,2}$ components with an r.m.s. of  $0.2$ and $0.1$ respectively.
Each galaxy is sheared with a uniform random $g_{1,2}$ value between $-0.04$ to $ 0.04$.
The half light radii of galaxy surface profiles before convolution with the PSF have a truncated Gaussian distribution, with mean $0.75$, r.m.s. $0.25$ and minimum $0.2$ in units of the half light radius of the PSF.
We use a Gaussian PSF with $FWHM=0.9''$, because the convolutional neural network was trained on stacked images with complex PSFs.
We assign a uniform random ellipticity to the psf with $\epsilon_{1,2}$ values between -0.025 to 0.025.

\subsection{\textsc{metacalibration} results}

\begin{figure} 
 \includegraphics[width=1\columnwidth]{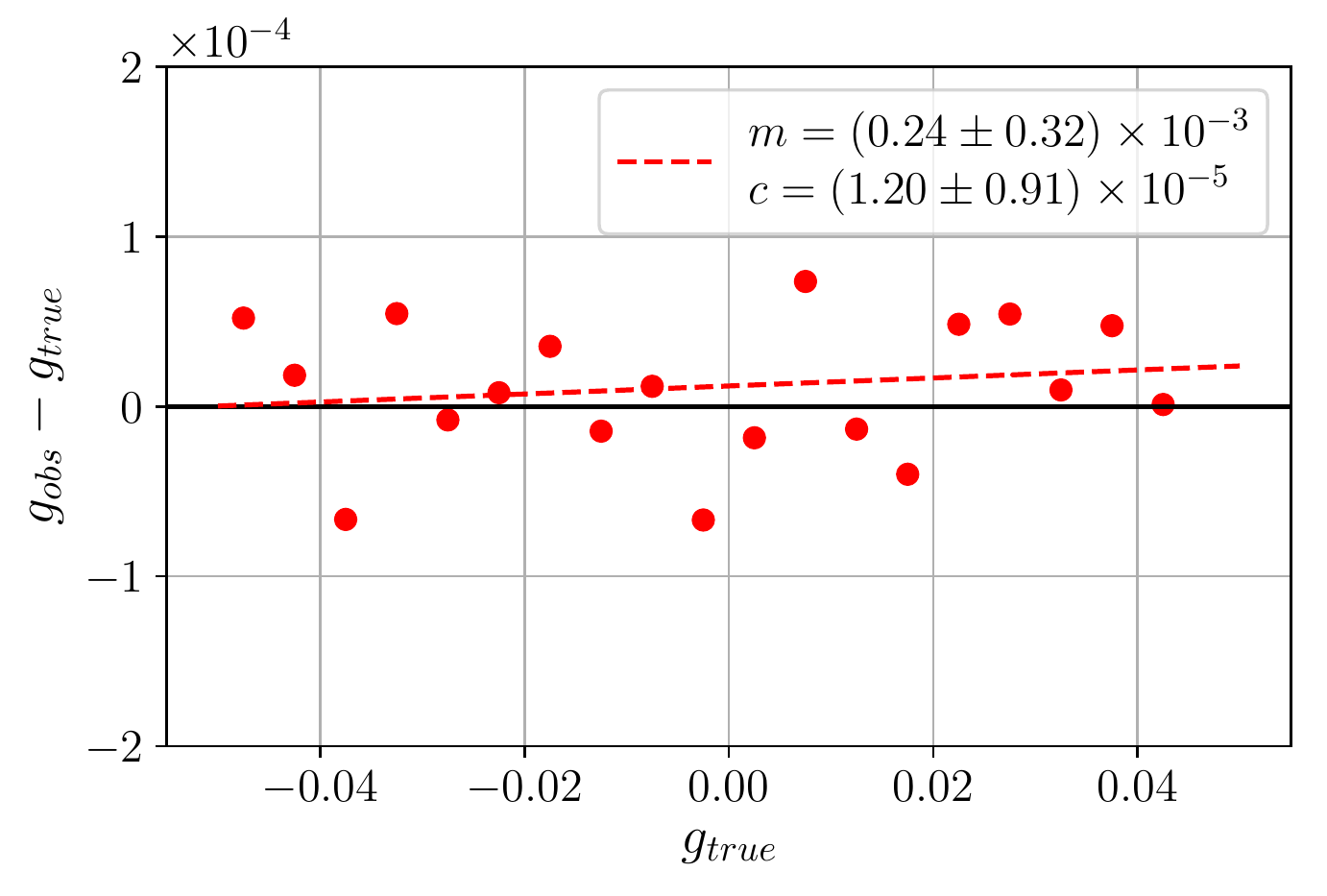}
    \caption{Multiplicative and additive bias of the metacalibrated CNN shear estimator. Measured from 350 million simulated galaxies.}
    \label{fig:mc}
\end{figure}

We implement the \textsc{metacalibration} procedure using CNN shape estimates building on the convolution and shear functions of galsim.
We use the CNN described previously, trained only on single band DES images and adopt an artificial shear value $\Delta g = 0.001$.
The convolution operations of \textsc{metacalibration}  dominate the runtime, approximately 40 ms per galaxy, therefore we use full test-time augmentation by averaging predictions on all 8 flipped and transposed combinations, as it was practically free.
The mean response of the CNN shape estimator on the simulation is around $0.7$, apparently smaller than in the observational data, showing that the simulation is sufficiently challenging.
Multiplicative and additive biases are calculated by fitting a linear function to the error of shear estimates depending on known shear simultaneously for 350 million galaxies.
We find that biases of the CNN estimator after \textsc{metacalibration} are the same in the two components, therefore we merge them.

The final multiplicative bias is $m=(0.24\pm 0.32) \times 10^{-3}$, well below the limit necessary for future large weak lensing surveys.
The result is very similar to the one obtained by \citep{sheldon_practical_2017} using their variable shear simulation.
The additive bias is also negligible, $c=(1.2\pm 0.9) \times 10^{-5}$.
The results are depicted with the fitted linear relationship and 20 aggregated points [Fig. \ref{fig:mc}].

\begin{figure} 
 \includegraphics[width=1\columnwidth]{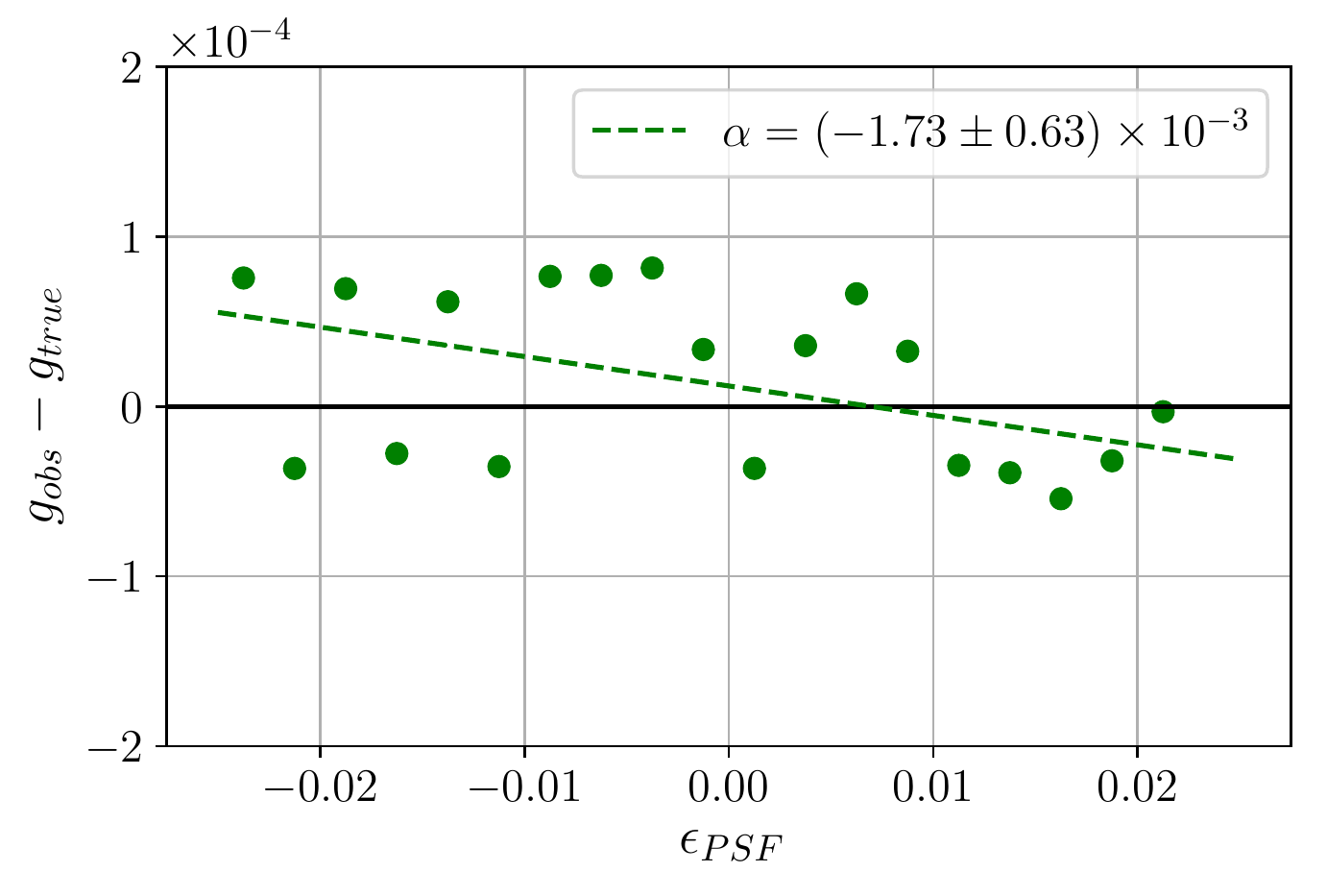}
    \caption{PSF leakage of the metacalibrated CNN shear estimator.}
    \label{fig:alpha}
\end{figure}

In \textsc{metacalibration}, PSF asymmetries are handled by PSF measurement and deconvolution instead of shape measurement algorithms, and  it was thoroughly tested on DES Y1 data by \cite{zuntz_dark_2018}.
They found that PSF measurement errors dominate the potential error introduced by deconvolution.
We also characterise the PSF leakage of shape measurement by fitting a linear function to shape estimate errors and PSF ellipticity.
We find that PSF leakage is an order of magnitude smaller than the effect of PSF measurement errors in \cite{zuntz_dark_2018} [Fig. \ref{fig:alpha}].

\begin{figure} 
 \includegraphics[width=1\columnwidth]{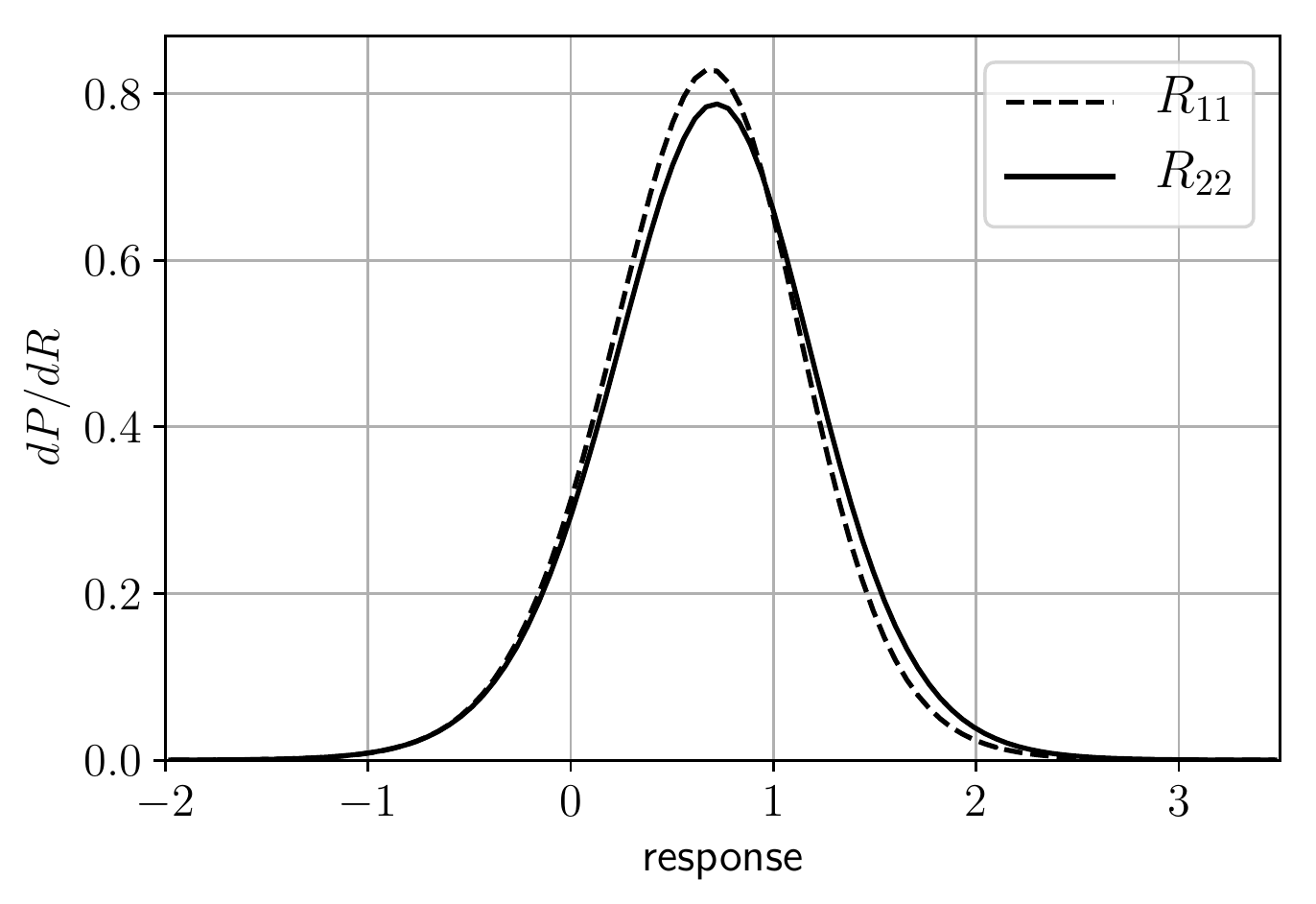}
    \caption{The distribution of the response values of the CNN. Mean response is larger for the diagonal ellipticity component.}
    \label{fig:reps}
\end{figure}

Interestingly, the response of the CNN is different for the two shape components, with a significantly larger response for the second, diagonal component [Fig. \ref{fig:reps}]. 
The difference is due to additional shearing operations and the application of fixnoise, because multiplicative biases of the two components are equal on raw simulated images.
However, we find no such difference when estimating shapes of these galaxies using adaptive moments \citep{bernstein_shapes_2002}.
Note that this peculiarity does not interfere with the \textsc{metacalibration}  procedure.

\section{Discussion}
\label{discuss}
We present a novel setup to create a training and testing dataset from only observational data using a wider but shallower survey to obtain postage stamp images of galaxies, and a narrow but deeper survey to measure the 'ground truth' shapes of galaxies.
Such a training and testing setup will also be possible to construct for any ongoing or planned weak lensing survey with dedicated deep fields such as Deep Drilling Field for LSST, Euclid Deep Fields, The Subaru Deep Field and WFIRST Deep Fields.
Space telescope data from Hubble and WJT could also be used to provide ``ground truth''  shapes.

%
We propose convolutional neural networks for galaxy shape estimation for weak lensing, and we show that a CNN is able to perform a shape measurement in $0.2$ milliseconds, more than $10000\times$ faster than a maximum likelihood forward fitting method.
Model fitting procedures are very resource intensive, which already warrant simplifications in current surveys \citep{conti_calibration_2017,zuntz_dark_2018}, and it will be a large challenge to tackle large scale, deep surveys in the near future.
CNNs offer a super-fast alternative, which could significantly facilitate data analysis for very large weak lensing surveys.

%
The CNN is able to match the precision of maximum likelihood shape estimates for very bright galaxies and significantly outperforms them on faint galaxies.
On the small and faint subset of a galaxies excluded from cosmology in the DES \textsc{im3shape} catalogue, where the maximum likelihood method struggles to provide reliable shape estimates which can be calibrated, the CNN is able to measure the shapes of this galaxies with much higher accuracy, potentially preventing the exclusion of some these galaxies for increased galaxy density.
The CNN is able to measure galaxy shapes using multi-band input images with no additional overhead compared to the single band case.
The precision of these multi-image shape estimates match the DES Y1 \textsc{metacalibration} catalogue at bright galaxies, and it is significantly higher at faint galaxies.
The CNN is not limited by simple surface brightness profiles, which enables it to significantly outperform the single Gaussian fit of DES Y1 \textsc{metacalibration} on blue galaxies, while the difference is much smaller on relaxed red galaxies.

%
Improved precision is worthless if systematics degrade, and we demonstrate that the CNN shape estimator can be nested in the \textsc{metacalibration} process, to yield shear estimates with a sufficiently small multiplicative bias for future large weak lensing surveys $m<10^{-3}$  with no significant PSF leakage.
%
The CNN reaches peak performance around $4 \times 10^5$ galaxies, which can be easily collected in future large surveys, and our result indicate that more data might not improve precision much further.
%

For our proof of concept study we use the easily accessible stacked images from DES DR1, however, weak lensing surveys generally evaluate multi-epoch data with a joint fit over single exposures \citep{miller_bayesian_2013, zuntz_dark_2018} as the interpolation of the PSF creates artefacts in stacked images which are problematic for shear estimation  \citep{miller_bayesian_2013}.
A joint prediction using multiple exposures is perfectly possible with CNNs too, as demonstrated by joint prediction on multiple band images.
The exploration of CNN shape estimators in extreme cases such as $\approx$ 1000 exposures per galaxy in LSST may provide further interesting results.

\textsc{metacalibration} reduces the signal to noise ratio of galaxies and therefore hampers the precision of shear estimates in exchange for robustly eliminating bias.
Our current implementation of \textsc{metacalibration}  dominates runtime, therefore in order to fully exploit the performance of CNNs this limitation needs to be tackled, possibly with a GPU accelerated  \textsc{metacalibration} implementation.
The results of \cite{tewes2019weak} indicate that a specifically tailored neural network could be able to produce shear estimates with small bias and no significant PSF leakage.
\cite{tewes2019weak} provide properties of the galaxy and the PSF as inputs to a densely connected neural network, and a CNN could also have access to all the necessary information and it could also be specifically trained to natively produce shear estimates with small bias and PSF leakage.
Such a shape estimator could fully realise the potential of CNNs both in terms of precision and speed.

Further tests with CNN shape estimators in simulations could improve our understanding of factors which contribute to the performance advantage of the CNN compared to a maximum likelihood model fitting approach.
The extraction of meaningful and interpretable knowledge from the inspection of a CNN could also improve our understanding of the problem itself  \citep{ribli2019improved}.
%
%
Finally, in our proposed scheme, galaxy shape estimation with CNNs cannot completely replace model fitting approaches, as the training procedure relies on high-quality shape measurements from a deeper survey which must be performed with conventional methods.

\section*{Acknowledgements}

We thank the anonymous referee, for the report which helped us significantly improve the manuscript, and we thank Joe Zuntz and Erin Sheldon for useful discussions and comments.
This work was partially supported by the National Research, Development and Innovation Office of Hungary via grant OTKA NN 129148 and the National Quantum Technologies Program. 

\bibliographystyle{mnras}

\bsp	
\label{lastpage}
\end{document}